# Design and Considerations of ADC0808 as Interleaved ADCs

**Rajiv Kumar[1], Rakesh Gupta[2]**

[1,2]Assistant Professor

[1]Department of Electronics & Communication Engineering

[2]Department of Electrical & Electronics Department

[1,2]U.P. Technical University, Lucknow, India

**Abstract—** Here in this paper we are presenting a digital system background technique for correcting the time, offset, error rate and gain mismatches in a time-interleaved analog-to-digital converter (ADC) system for N-channel communication using 8-bit ADC0808 IC's. A time-interleaved A–D converter (ADC) system is an effective way to implement a high-sampling-rate ADC with relatively slow circuits. This paper analyzes the benefits and derives an upper band on the performance by considering kT/C noise and slewing requirement of the circuit driving the system. In the system, several channel ADCs operate at interleaved sampling times as if they were effectively a single ADC operating at a much higher sampling rate. A timing mismatch calibration technique is proposedthat covers linear and nonlinear channel mismatches, unifies, and extends the channel models. A novel foreground channel mismatch identification method has been developed, which can be used to fully characterize dynamic linear mismatches. A background identification method provides accurate timing mismatch estimates.

## I.     INTRODUCTION

An A/D converter is a device that converts a continuous physical quantity (usually voltage) to a digital number that represents the quantity's amplitude. The conversion involves quantization of the input, so it necessarily introduces a small amount of error. Instead of doing a single conversion, an ADC often performs the conversions periodically. An ADC is defined by its bandwidth and its signal to noise ratio. The actual bandwidth of an ADC is characterized primarily by its sampling rate and to a lesser extent by how it handles errors such as aliasing. The dynamic range of an ADC is influenced by many factors, including the resolution (the number of output levels it can quantize a signal to), linearity and accuracy (how well the quantization levels match the true analog signal) and jitter (small timing errors that introduce additional noise). The dynamic range of an ADC is often summarized in terms of its effective number of bits (ENOB), the number of bits of each measure it returns that are on average not noise. An ideal ADC has an ENOB equal to its resolution. ADCs are chosen to match the bandwidth and required signal to noise ratio of the signal to be quantized. If an ADC operates at a sampling rate greater than twice the bandwidth of the signal, then perfect reconstruction is possible given an ideal ADC and neglecting quantization error. The presence of quantization error limits the dynamic range of even an ideal ADC.

### A. Introduction to Interleaving  process

The concept of time interleaving was originally proposed as a means of increasing the speed of analog to digital converters [1]. Because of ever-increasing demand for higher data rate and larger bandwidth, modern digital communication systems depend onanalog-to-digital converters (ADCs) operating at faster speed and providing higher resolution. Although it is difficult to scale ADCs to satisfy these performance requirements while maintaining low production costs. The time-interleaving multiple ADCs is a well-known approach to increase the sampling rate [1],[2] while keeping hardware costs at bay. A–D converters (ADCs) incorporated in such instruments have to operate at a very high sampling rate. This paper studies theoretical issues of a time-interleaved ADC system where several channel ADCs operate at interleaved sampling times as if they were effectively a single ADC operating at a much higher sampling rate [3]–[7]. Fig. 1 shows such an ADC system where each M-channel ADCs ($ADC_0$, $ADC_1$…… $ADC_{M-1}$) operates with one of M-phase clocks ($CK_0$, $CK_1$……$CK_{M-1}$) respectively. The sampling rate of the ADC as a whole is Mtimes the channel sampling rate. This time-interleaved ADC system is an effective way to implement a high-sampling-rate ADC with relatively slow circuits, and is widely used. Ideally, characteristics of channel ADCs should be identical and clock skew should be zero.

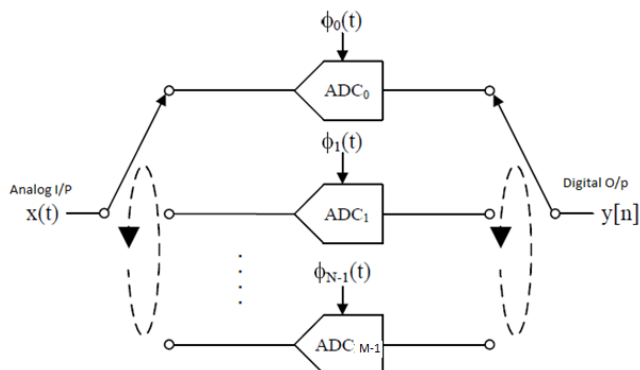

Fig.1. Block diagram of the time-interleaved ADCs for M-channel.

However, in reality there are mismatches such as offset, gain mismatches among channel ADCs as well as timing skew of the clocks distributed to them, which cause so-called pattern noise and significantly degrade S/N ratio of theADC system as a whole. Once the mismatch errors are estimated, we have two ways toeliminatetheerror effects.Onewayistoadjustthe sampling clockforeachADC [8], whichmayincrease herandomjitter ofeachcontrolledclock.Anotherwayistodothe interpolation to achieve the correct values at the ideal times [9]. The second approach is attractive because it can be done with the required accuracy using digital signal- processing circuits, which are portable and will benefitfrom evolving scaled CMOS technologies [10].

### B.   Time Interleaving

For very-high-speed applications, time interleaving increases the overall sampling speed of a system by operating two or more data converters in parallel. This sounds reasonable and straightforward but actually requires much more effort than just paralleling two ADCs. Before discussing this arrangement in detail, compare the sampling rate of a time-interleaved system with that of a single converter. As a rule of thumb, operating M number of ADCs in parallel increases the system's sampling rate by approximately a factor of M. Thus, the sampling (clock) frequency for an interleaved system that hosts M ADCs can be described as follows:

$$F_{ADC\_clock} \leq \sum_{n=1}^{n=M-1}(fclock\_adc)^2$$

The simplified block diagram in Fig.2 illustrates a single-channel, time-interleaved DAQ system in which two ADCs double the system's sampling rate. This rate ($F_{ADC\_CLK}$) is a clock signal at twice the rate of $f_{CLK1}=f_{CLK2}$. Because $f_{CLK1}$ is delayed with respect to $f_{CLK2}$ by the period of $F_{ADC\_CLK}$, the two ADCs sample the analoginput signals alternately and result in producing an overall sample rate equal to $F_{ADC\_CLK}$. Each converter operates at half the sampling frequency.

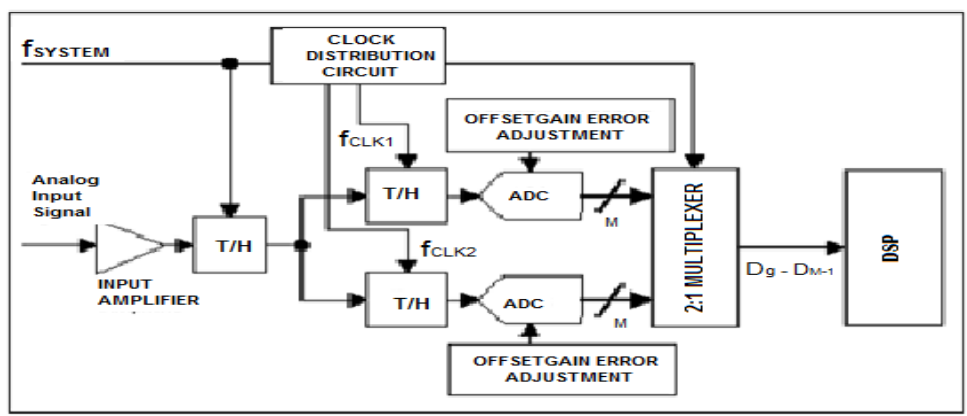

Fig. 2 Time interleaving

## C.   Timing Mismatch

A number of mismatch components displace the interleaved sampling instants from their ideal position. For two channels, the sampling clock path consists of a voltage-controlled oscillator (VCO), a buffer, a bootstrap circuit and a differential sampling network. Fig. 3 conceptually illustrates this path, including only the relevant sections of the bootstrap circuit [5].

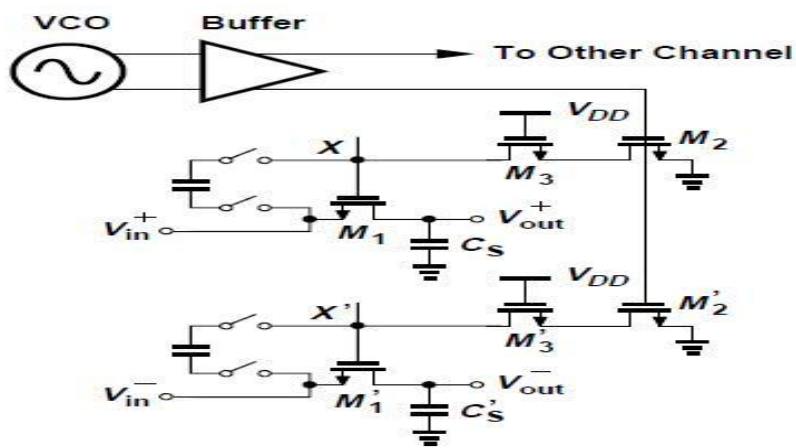

Fig. 3Typical clock path in a two-channel ADC

Here, transistor $M_2$ turns off the main sampling switch $M_1$ and $M_3$ shields M2 as $V_X$ rises above the supply voltage [11]. The other phase of the clock drives the second channel. The timing error in this case arises from the departure of the VCO duty cycle from 50% and the mismatches in the clock path. In addition to the asymmetries in the VCO and its buffer, the bootstrap devices introduce significant timing errors. This can be seen from Fig. 3 by noting that (a) the branch consisting of $M_2$, $M_3$, the total capacitance at node X, and $M_1$ contributes mismatches and (b) the standard deviation of these mismatches is multiplied by 2 because the two differential interleaved ADCs incorporate *four* such branches. For a large number of interleaved channels, the clock path becomes more complex and hence prone to larger mismatches. Fig. 4 shows an example for four channels. A divide-by-two circuits generates quadrature phases of the clock but also contributes additional mismatches.

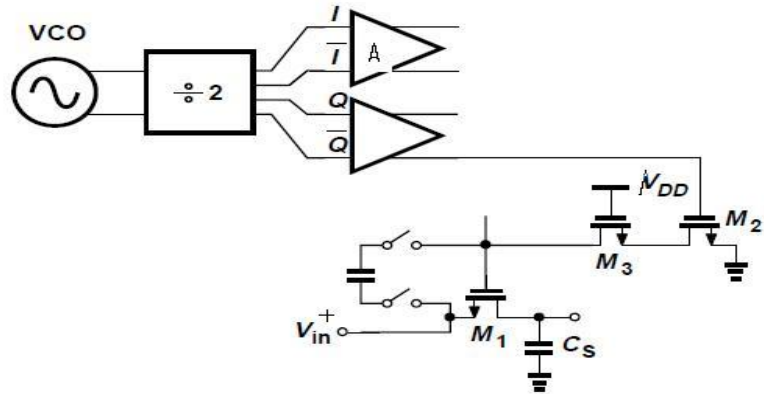

Fig. 4 Clock path for interleaving by a factor of 4 or higher

It is possible to remove the phase mismatches caused by the VCO and the divider through the use of gating [12]. Consider the two-channel where "predictive" waveforms $V_{even}$ and $V_{odd}$ gate the master clock such that its falling edges are alternately applied to each channel [12]. So long as each pulse on $V_{even}$ and $V_{odd}$ encloses the falling edge of $Ck_{master}$, the exact position or width of these pulses is immaterial. This scheme is insensitive to the duty cycle distortion of $Ck_{master}$ and other mismatches preceding switches $S_1$ and $S_2$ i.e. the timing error arises from only these switches and the sampling transistors, $M_1$ and $M_2$. The narrow predictive waveforms may suggest insufficient acquisition or hold time. This issue is resolved by means of the two channels sampling commands, $Ck_1$ and Ck2, are derived from $Ck_{master}$ . A pulse on $V_{odd}$ routes the falling edge of $Ck_{master}$ to $Ck_1$ and the rising edge of $Ck_{master}$ to $Ck_2$. After this pulse subsides, $Ck_1$ and $Ck_2$ retain their values (as in dynamic logic) until the next pulse on $V_{even}$ arrives. As a result, $Ck_1$ and $Ck_2$ provide about 50% of the clock cycle for acquisition. The critical timing mismatches still stem from only $S_1$-$S_2$ and $M_1$-$M_2$ pairs. This scheme can be generalized to more than two channels so as to remove errors due to frequency dividers, phase interpolators, etc.

### D.  *Offset and Gain Errors*

The channel-to-channel matching of offset and gain in separate ADCs is not trimmed, so gain and offset mismatches between ADCs are parameters of concern in a time-interleaved system. If one ADC shows an offset and the other a gain error, the digitized signal represents not only the original input signal but also an undesired error in the digital domain. An offset discrepancy causes a signal phase shift in the digitized signal, and gain mismatches show up as differences in signal amplitude.

## II. DESIGN AND ANALYSIS

The additional channel ADC can be triggered with the same clock signal asone of the other channel ADCs. Hence, in the ideal case, i.e., without any mismatches, ittakes samples at the same time instants as one of the other channel ADCs. The principleisillustrated in Fig. 5 and the corresponding timing is shown in Fig. 6.The method ismuch more robust against correlation between the switching sequence and the input signal and compared to purely digital identification methods, it requires much shorter observationtimes.The reference channel can be any channel in the time-interleaved ADC i.e. it does nothave to be calibrated. However, a calibrated reference ADC can improve the estimationresults.

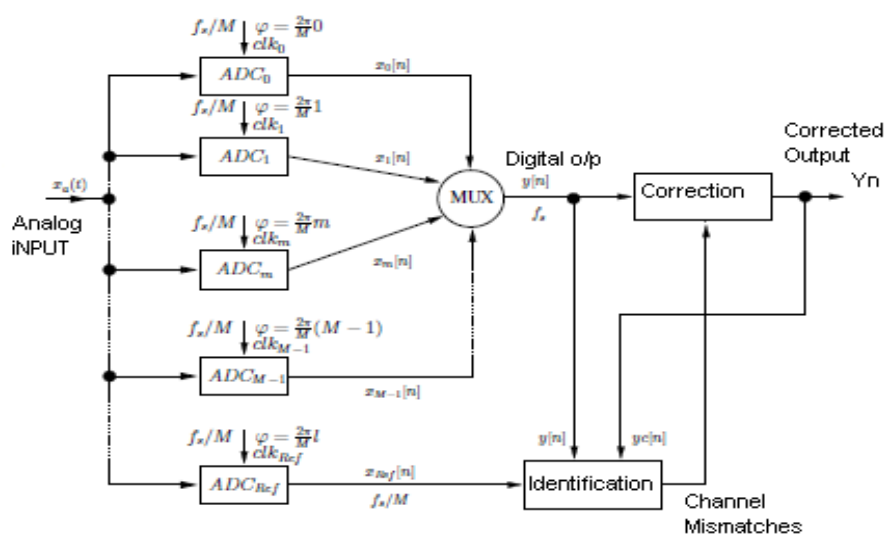

Fig. 5 Architecture of a time-interleaved ADC with a reference channel ADC

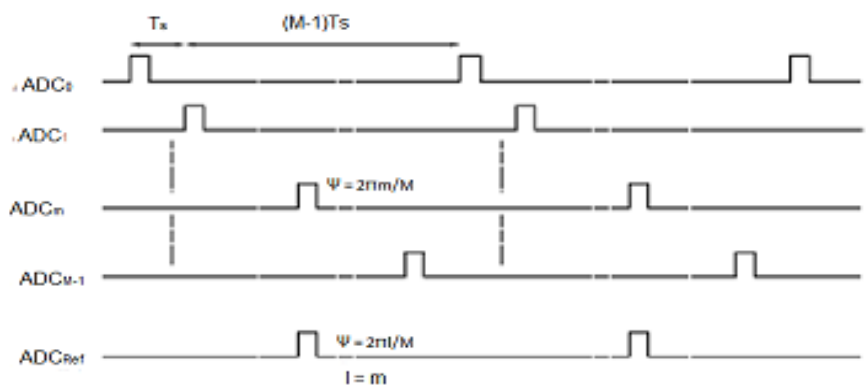

Fig. 6Timing diagram for the identification using reference channel ADC.In the reference channel has the same clock signal as the M[th]channel ADC

### A. Analysis ofGain Mismatch Identification

In order to identify gain mismatches, we compare the power of the reference channel and the selected channel ADC outputs. When we first compensate offset mismatches, any deviation of the channel power has to be caused by gain mismatches, since timing mismatches do not cause power variations among the channels. Thus, the gain mismatches are given by

$$g_m = \sqrt{\frac{\sum_{k=0}^{K-1} y^2 m(k)}{\sum_{K=0}^{K-1} y^2 Ref(k)}}$$

To show the performance of the algorithm, we simulate a time-interleaved ADC with eightChannels (M = 8) and an additional reference channel. Except that the considered channel ADC has gain and timing mismatches, the configuration and the input signal specificationis identical to the offset identification simulations. In fig. 7, the uncompensated curve is saturated because the timing mismatch limits the performance and a decreasing gain mismatch standard deviation has no influence on the performance. Hence, the gain identification is only limited by the timing mismatches.

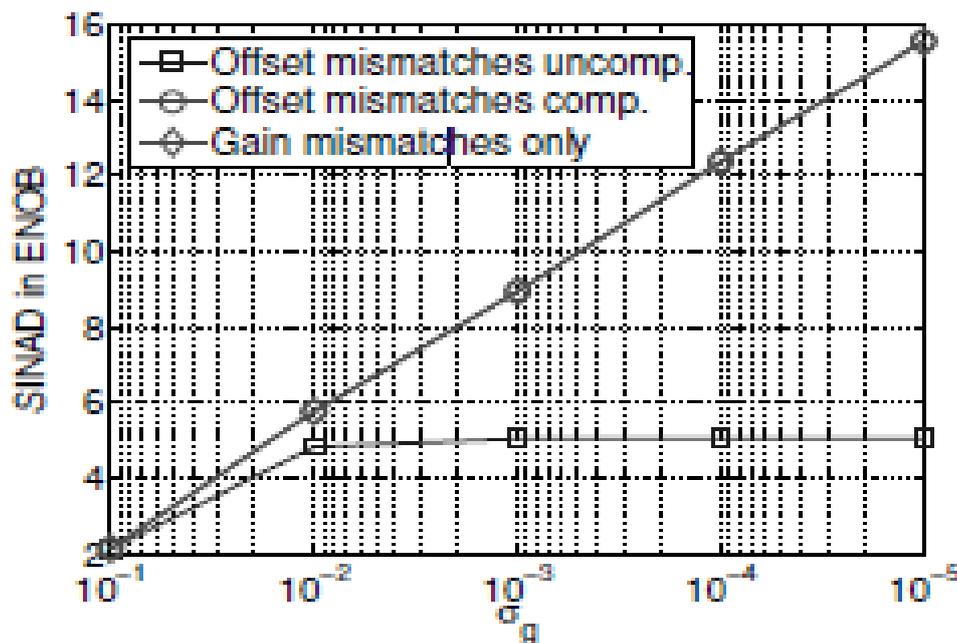

Fig. 7The figure shows the worst-case identification quality for different gain mismatch standard deviations and constant offset mismatches

## B.   *Analysis of Timing Mismatch Identification*
## C.

The problem of the timing mismatch identification algorithm presented in [12] is its sensitivity to the input signal frequency distribution. With the reference channel approach we can improve the accuracy as well as the identification stability of the algorithm. For the method presented in [12], the reference is the average over all channel ADCs. In our case, we can use the additional channel ADC as reference. The principle is explained in Fig. 8. Without any mismatches, the amplitude difference between the reference channel output at time$(kM + m)Ts$ and the proceeding channel ADC output at time $(kM + m-1)Ts$equals the difference of the current channel ADC output and the proceeding channel ADC output. As soon as we have timing mismatches, these differences diverge. For negative timing mismatches, i.e., the clock signal is delayed, the magnitude difference $y_m[k] - y_{m-1}[k]$ is on average larger than the magnitude difference of $y_{Ref}[k]$  - $y_{m-1}[k]$ and vice versa for positive timing mismatches.

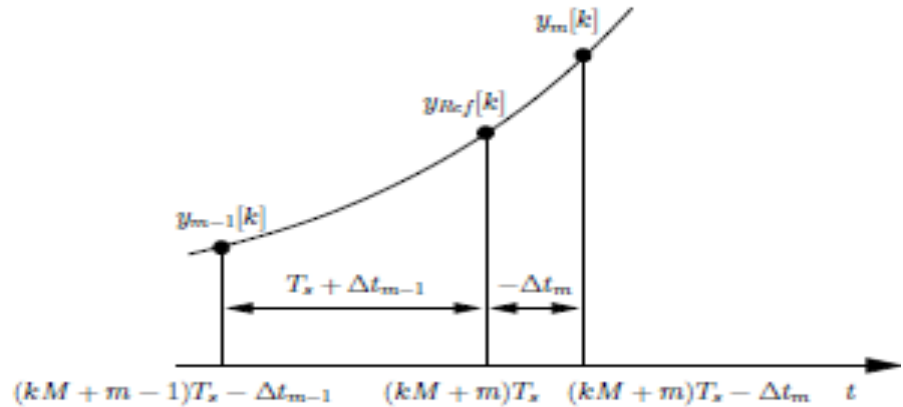

Fig. 8 The figure illustrates the identification of timing mismatches with a referencechannel ADC

In Fig. 8, we have shown an example with a positive slope.For negative slopes we obtain the same results with opposite signs. Thus we can write

$$\frac{[\text{yRef}(k) - ym-1(k)]}{Ts + \Delta tm-1} = \frac{[ym(k) - ym-1(k)]}{Ts + \Delta tm-1 - \Delta tm} \dots\dots\dots(a)$$

Where eq. (a) is only exact when the $m^{th}$ channel ADC has no timing mismatches i.e. $\Delta tm= 0$, Otherwise, eq.(a) describes the best local linear approximation to the signalevolution. In this case, we can use the following relation to determine the timing mismatchesinstesd of eq. (a).

$$\text{E}\{\frac{[\text{yRef}(k) - ym-1(k)]}{Ts + \Delta tm-1}\} = \text{E}\{\frac{[ym(k) - ym-1(k)]}{Ts + \Delta tm-1 - \Delta tm}\} \dots\dots\dots (b)$$

$$\frac{\Delta tm}{Ts + \Delta tm-1} = 1 - \text{E}\{\frac{E\{|ym - ym-1|\}}{E\{|yRef - ym-1|\}}\} \dots\dots\dots (c)$$

Further we can use the relation

$$\frac{\Delta tm}{Ts + \Delta tm-1} \cong \frac{\Delta tm}{Ts} \cong rm \qquad\qquad \dots\dots\dots(d)$$

Where

$$rm = 1 - \text{E}\{\frac{E\{|ym - ym-1|\}}{E\{|yRef - ym-1|\}}\}$$

$t_m$= time mismatches after $m^{th}$time of A/D conversion
$t_{m-1}$ = time mismatches after $(m-1)^{th}$ time  of A/D conversion
$T_s$ = sampling period
$r_m$= relative timing mismatches
$k$ = number of samples
$y_m[k] - y_{m-1}[k]$ = magnitude difference

The basic assumption of the method is that, for a negative timing mismatch, we obtaina larger difference between the selected channel ADC and its predecessor compared to thedifference of the reference channel ADC and its predecessor.

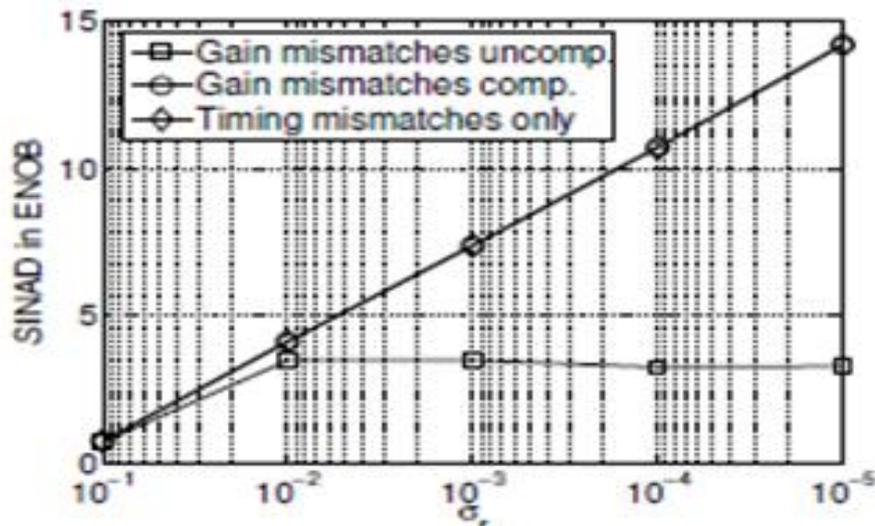

Fig. 9Worst SINAD for the gain mismatches identification method as a function of the timing mismatch standard deviation for the compensated and the uncompensated case. With this method, the gain mismatch identification is always accurate

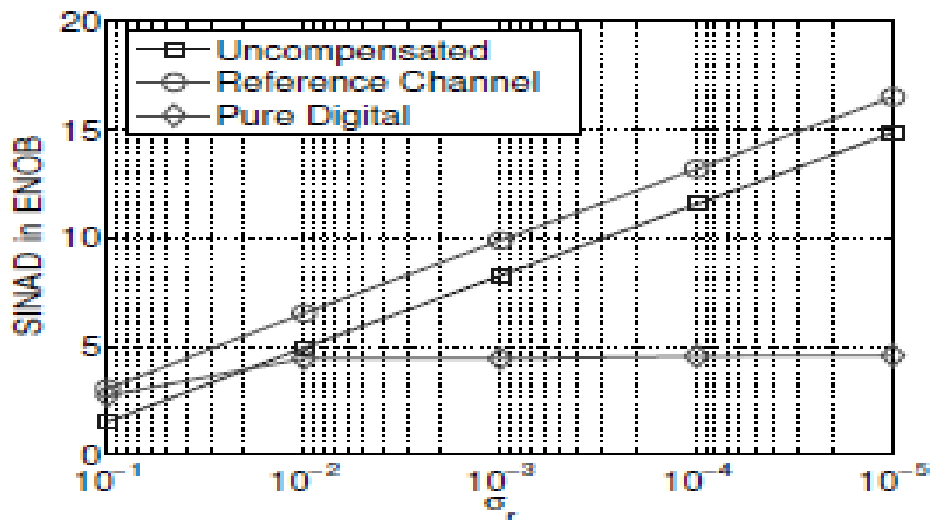

Fig. 10 The figure shows the average identification quality for different timing mismatch standard deviations

### D.  Time-Interleaved ADC Model:

Due to its wide range of toolboxes of MATLAB, we decided to use MATLAB as a simulation environment, which allows a fast and flexible programming of the required functionality. The simulation model as shown in fig. 11 should include the most important signal processing error sourcesof a time-interleaved ADC. The three main error sources degrading the spurious free dynamic range (SFDR) and the signal-to-noise and distortion ratio (SINAD) of a timeinterleavedADC are offset, gain and timing mismatches. Gain mismatches are the differences among the gains and offset mismatches are the differences among the offsets of the channel ADCs. Timing mismatches are the deterministic deviations between the nominal sampling time and the real sampling time for each channel. Another important error source is timing jitter, which is, however, not restricted to time-interleaved ADCs. In contrast to the timing mismatch, it is the stochastic deviation from the nominal sampling period for each sample [13]. From multi-rate systems analyses we also find that bandwidth mismatches lower the SINAD and the SFDR [14]. Conventional ADCs have static and dynamic nonlinearities, which have to be considered in time-interleaved ADCs as well.

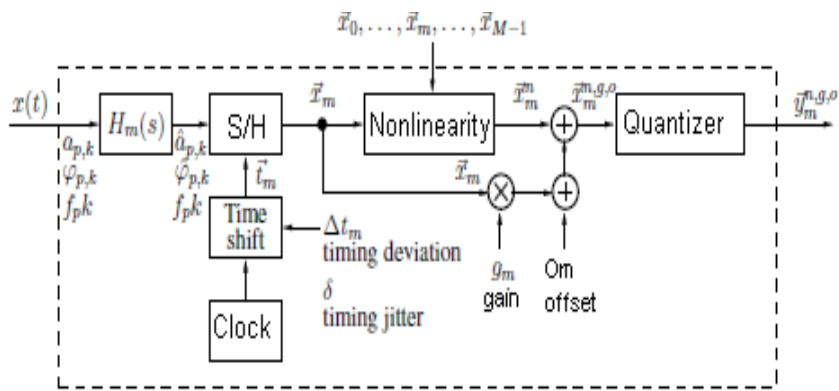

Fig. 11 Behavioral simulation model of the l[th] channel ADC.

This signal representation spans a sufficiently large signal space that allows us to simulate broad class of possible input signals (including non-harmonic signals) and can easily be applied to continuous-time filter operations. Moreover, the simulation can imitate the sampling process with high calculation precision through direct calculation of the continuous time since functions for any specific continuous-time instant t.

### III. SIMULATION RESULTS

In the following we will present our simulation results of the most important error sources, which show the abilities of the environment. In Fig. 12 we see the output spectrum of time-interleaved ADC consisting of four channel ADCs with offset, gain, and timing mismatches. Additionally, the time-interleaved ADC produces timing jitter. Beside the input signal, i.e., a sine wave given as x(t) = sin(2πf₀t), we see additional spurious linesin the spectrum. The spectral lines appear at kΩ/M for offset mismatches and at $\pm\Omega_0$ +kΩ$_s$/M for timing and gain mismatches. Although we use a 10-bit resolution and 1024samples for this simulation, we see a noise floor, which is considerably above 89 dBc of an ideal 10-bit ADC. This is due to the timing jitter, which increases the noise floor.

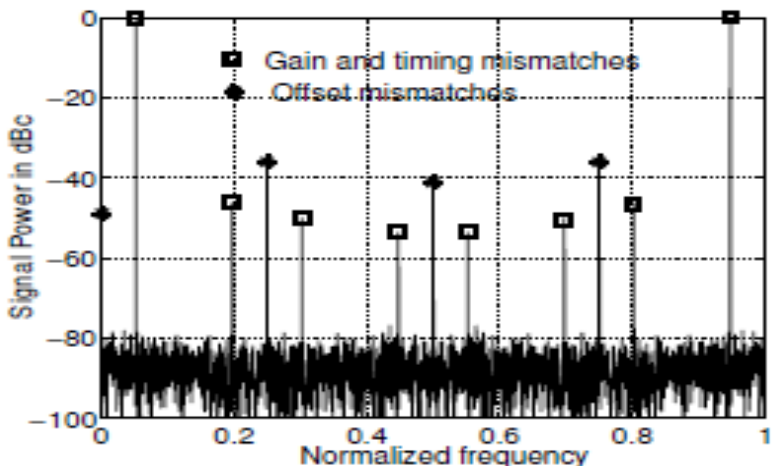

Fig. 12Simulation with offset, gain and timing mismatches

Another important error occurring in time-interleaved ADCs is the bandwidth mismatch. In order to simulate this error, we assume low-pass characteristics for the input filters and an ideal cut-off frequency (-3 dB), which is five times higher than the sampling frequency. Each input filter has a cut-off frequency that deviates randomly from the ideal one. Thesedeviations are assumed to be Gaussian distributed with a standard deviation of 0.1. Thetransfer functions of the described filters are shown in Fig. 13. We notice additionalspectral lines similar to the case with static gain and timing mismatches.

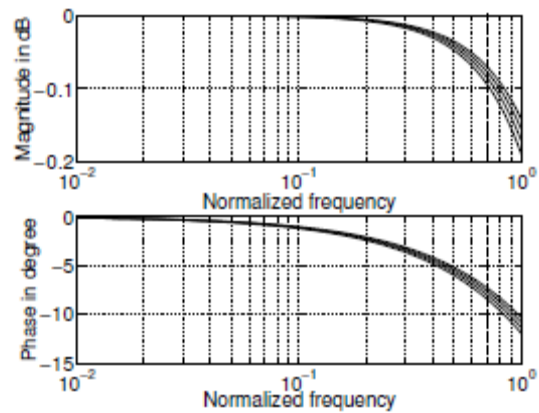

Fig. 13Input filters used for demonstrating the bandwidth mismatch effect. Thedashed line indicates half the sampling frequency

## IV. CONCLUSION

We have investigated time-domain methods for channel mismatch identification. As simulations and calculations have shown, purely digital methods work well as long as we donot have a correlation between the input signal and the switching sequence. However, weonly have investigated the methods for individual mismatches (offset, gain, and timing mismatches) and have not considered the identification of combined mismatches, i.e. caseswhere we identify one mismatch, e.g. offset mismatch, in the presence of other mismatches, e.g. gain and timing mismatches. For mismatch identification with a reference channel we have also investigated cases for combined channel mismatches. The results are very promising, since the identification of individual mismatches, e.g. offset mismatches, does not seem to be influenced by other mismatches e.g. gain and timing mismatches.

## ABOUT THE AUTHORS

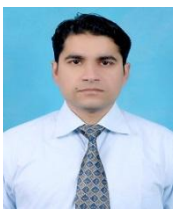

Rajiv Kumar received the B.Tech degree in Electronics and Communication Engineering from U.P. Technical University, Lucknow, India in 2008. He received M.Tech  degree from Subharti University, Meerut, India in 2013. During his post-graduation work, he researched on various real time RF signals, Microwaves and real time Antenna Design.  He is currently working as Assistant Professor in Roorkee Engineering and Management Technology Institute ,Uttar Pradesh, India.

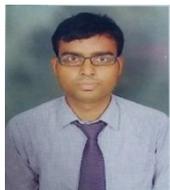

Rakesh Gupta was born in Ghazipur, Uttar Pradesh, India. He received the B.E. Degree (Electrical and Electronic Engineering) from Anna University, Chennai, India. During his Graduate Studies he held several workshop and summer training program. He is Currently an Assistant Professor  withRoorkee Engineering And Management Technology  Institute, Uttar Pradesh, India . He hasresearch interest in low power system design , Robotic ,Embedded system .